# Pressure-Induced Superconductivity in BiS$_2$-based EuFBiS$_2$

Kouji Suzuki[1,2], Masashi Tanaka[1,*], Saleem J. Denholme[1,†], Masaya Fujioka[1,‡],

Takahide Yamaguchi[1,2], Hiroyuki Takeya[1], and Yoshihiko Takano[1,2]

[1]*National Institute for Materials Science, 1-2-1 Sengen, Tsukuba, Ibaraki 305-0047, Japan*

[2]*Graduate School of Pure and Applied Sciences, University of Tsukuba, 1-1-1 Tennodai, Tsukuba, Ibaraki 305-8577, Japan*

**Abstract**

We measured the electrical resistivity of the BiS$_2$-based compound EuFBiS$_2$ under high pressure. Polycrystalline EuFBiS$_2$ shows insulator-metal transition and pressure-induced superconductivity above 0.7 GPa. The superconducting transition temperature increases with increasing applied pressure and shows a maximum value around 8.6 K at 1.8 GPa.

*E-mail: Tanaka.Masashi@nims.go.jp

[†]Present address: *Department of Applied Physics, Tokyo University of Science, 6-3-1 Niijuku, Katsushika, Tokyo 125-8585, Japan*

[‡]Present address: *Research Institute for Electronic Science, Hokkaido University, N20W10, Kita-ku, Sapporo, Hokkaido 001-0020, Japan*

Since the discovery of superconductivity in $Bi_4O_4(SO_4)_{1-x}Bi_2S_4$,[1] much attention has been paid to developing $BiS_2$-based layered superconductors. $BiS_2$-based compounds have a layered structure composed of an alternate stacking of superconducting and blocking layers, which is a common feature of cuprate or Fe-based superconductors.[2,3] The typical $BiS_2$-based compound $LaOBiS_2$ shows superconductivity by O substitution with F in the blocking layer,[4] which supplies electron carriers into the $BiS_2$ superconducting layer. This led to considerable advances in the $BiS_2$-based superconducting family within a few years.[5-18] Interestingly, many $BiS_2$-based superconductors are sensitive to external pressure.[19-22] The $T_c$ enhancement has been attributed to the local structure sensitivity of the $BiS_2$-based superconducting materials.[23]

Recently, it has been reported that the $BiS_2$-based compound $EuFBiS_2$ shows superconductivity even without any chemical doping.[24-26] A charge-density-wave (CDW)-like behavior was simultaneously observed in its electrical resistivity.[24] It is necessary to have more information about its physical properties by using other probes. In this study, we demonstrate the electrical resistivity measurements of $EuFBiS_2$ under high pressure. It is found that the semiconducting sample shows insulator-metal transition and superconductivity is induced by only applying pressure.

Polycrystalline samples of $EuFBiS_2$ were prepared by a solid-state reaction. Powders of $EuS$, $BiF_3$, and $Bi_2S_3$ were weighed with the nominal composition of $EuFBiS_2$. The mixture was well-ground and pressed into pellets. The pellets were sealed in an evacuated quartz tube and then heated at 700 °C for 20 h. X-ray diffraction (XRD) measurement with Cu K$\alpha$ radiation was carried out using Mini Flex 600 (RIGAKU). The compositional ratio was analyzed by energy dispersive X-ray spectroscopy (EDX) using JSM-6010LA (JEOL). The electrical resistivity measurement was performed in the temperature range from 2 to 300 K using a physical property measurement system (PPMS, Quantum Design) by a four-probe method. A piston-cylinder-type high-pressure cell was used for applying hydrostatic pressure to the sample. Fluorinert 70/77 was employed as a pressure-transmitting medium. All the pressure values were estimated from the $T_c$ of a Pb manometer, although the actual pressure around room temperature includes underestimation.

The obtained sample is the single phase of EuFBiS$_2$, and the diffraction peaks can be indexed on the basis of a tetragonal unit cell with the lattice parameters $a$ = 4.0478(7) Å and $c$ = 13.520(3) Å, as shown in Fig. 1. The compositional ratio is estimated to be Eu: F: Bi: S = 1: 0.9(4): 0.9(5): 1.7(7) from EDX analysis, which is in good agreement with the nominal composition of EuFBiS$_2$ within the error.

The electrical resistivity at ambient pressure shows semiconducting behavior with a broad hump at around 230 K [Fig. 2(a)]. The resistivity increases with decreasing temperature, and is insulating in the temperature range below 180 K. The behavior drastically changes upon compression. The insulating behavior turns to metallic with applying pressure. This insulator-metal transition occurred at the pressure between 0.3 and 0.7 GPa. And also, the superconducting transition clearly appears at the pressure of 0.7 GPa as shown in Fig. 2(b). The onset superconducting transition temperature ($T_c^{onset}$) and zero-resistivity temperature ($T_c^{zero}$) are 5.2 and 2.1 K, respectively, at a pressure of 0.7 GPa. As shown in the lower panel of Fig. 2(c), both $T_c^{onset}$ and $T_c^{zero}$ exhibit a bell-shaped curve with the maximum of 8.6 K at 1.8 GPa against the applied pressure.

The broad hump at around 230 K at ambient pressure was gradually suppressed upon compression, and the peak temperature of the hump ($T_{hump}$) decreased with increasing pressure, as shown in the upper panel of Fig. 2(c). It is interesting to note that there is a correlation between the hump suppression and superconducting appearance. When the hump is suppressed completely, $T_c$ reaches its highest value at 1.8 GPa.

Figure 3(a) shows the electrical resistivity at a pressure of 1.8 GPa under magnetic fields up to 3.5 T. The superconducting transition was suppressed with increasing magnetic field. The magnetic field dependences of $T_c^{onset}$ and $T_c^{zero}$ are shown in Fig. 3(b). The upper critical field ($H_{c2}$) and irreversible field ($H_{irr}$) were estimated to be 3.0 and 1.5 T, respectively.

It has been reported that the $T_c$ of $RE$(O,F)BiS$_2$ ($RE$ = La, Ce, Pr, Nd, Yb) series suddenly increases with increasing pressure.[19-22] In the case of EuFBiS$_2$, however, $T_c$ increases gradually up to its maximum with increasing pressure. The gradual increases in $T_c$ may evoke a correlation between $T_c$ and its electron density of states at the Fermi level. There is a possibility that the europium valence changes with applied pressure. If the europium valence changes from +2 to +3 with applied pressure, more carriers are doped into BiS$_2$ layers. The superconducting mechanism of EuFBiS$_2$ still attracts considerable interest. Further investigations are required to understand the intrinsic properties of superconductivity of EuFBiS$_2$, for example, structural analysis, magnetic property and specific heat measurements under pressure.

In conclusion, the superconducting transition in EuFBiS$_2$ was induced by only applying pressure. The electrical resistivity at ambient pressure showed semiconducting behavior with a broad hump at around 230 K. EuFBiS$_2$ showed insulator-metal transition and the superconductivity was observed at a pressure of 0.7 GPa. $T_c$ increased gradually with increasing pressure and showed a maximum of 8.6 K at a pressure of 1.8 GPa.

**Acknowledgement**

This work was partially supported by the Advanced Low Carbon Technology R&D Program (ALCA) of the Japan Science and Technology Agency.

**Figure captions**

Fig. 1 (Color online) XRD pattern of the polycrystalline sample of EuFBiS$_2$. Schematic illustration of the inset shows the crystal structure of EuFBiS$_2$.

Fig. 2 (Color online) (a) Temperature dependence of electrical resistivity for EuFBiS$_2$ under various pressures. The arrows indicate the peak temperature of the hump structure ($T_{\mathrm{hump}}$). (b) Enlargement scale of the superconducting transition. $T_{\mathrm{c}}^{\mathrm{onset}}$ was determined to the temperature at 95% of the resistivity in its normal conduction state. (c) Pressure dependences of $T_{\mathrm{hump}}$, $T_{\mathrm{c}}^{\mathrm{onset}}$ and $T_{\mathrm{c}}^{\mathrm{zero}}$.

Fig. 3 (Color online) (a) Temperature dependence of electrical resistivity under various magnetic fields at a pressure of 1.8 GPa. (b) Magnetic field dependences of $T_{\mathrm{c}}^{\mathrm{onset}}$ and $T_{\mathrm{c}}^{\mathrm{zero}}$. The dotted line in $H_{\mathrm{c2}}$ corresponds to the estimation from the Werthamer–Helfand–Hohenberg (WHH) approximation for the type-II superconductor in a dirty limit. The dashed line in $H_{\mathrm{irr}}$ is the linear extrapolation of $T_{\mathrm{c}}^{\mathrm{zero}}$.

**Figures**

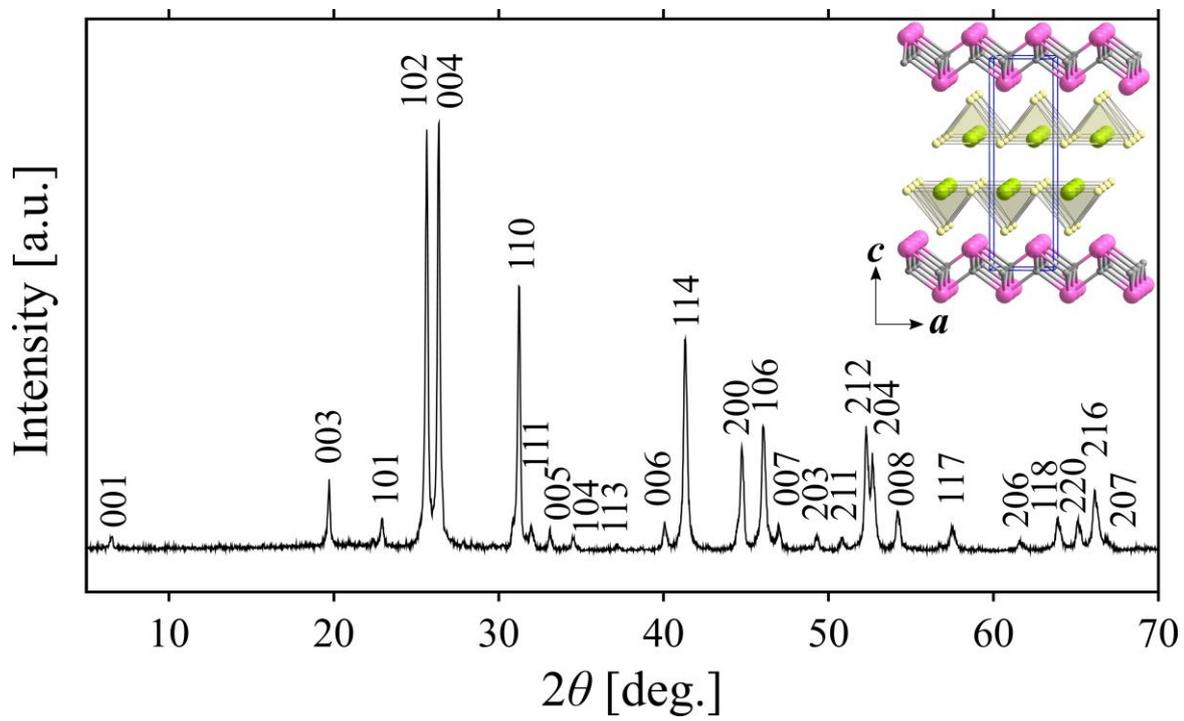

Figure 1

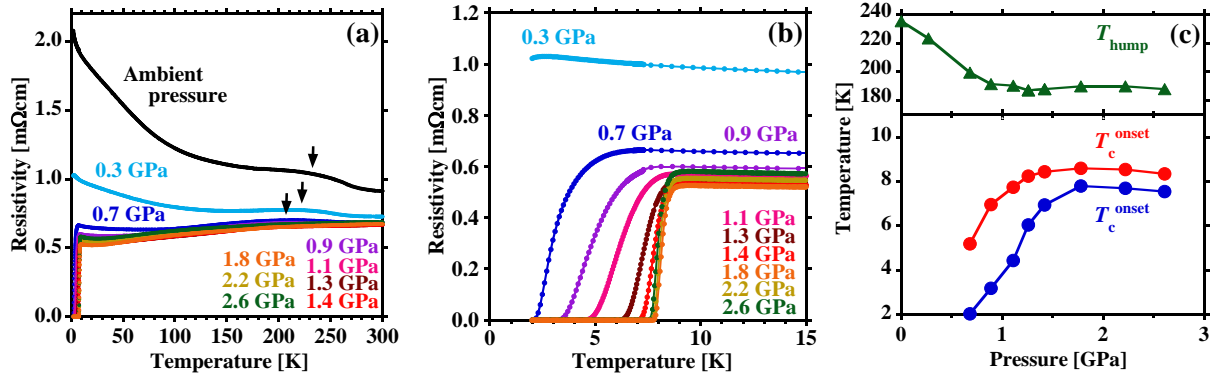

Figure 2

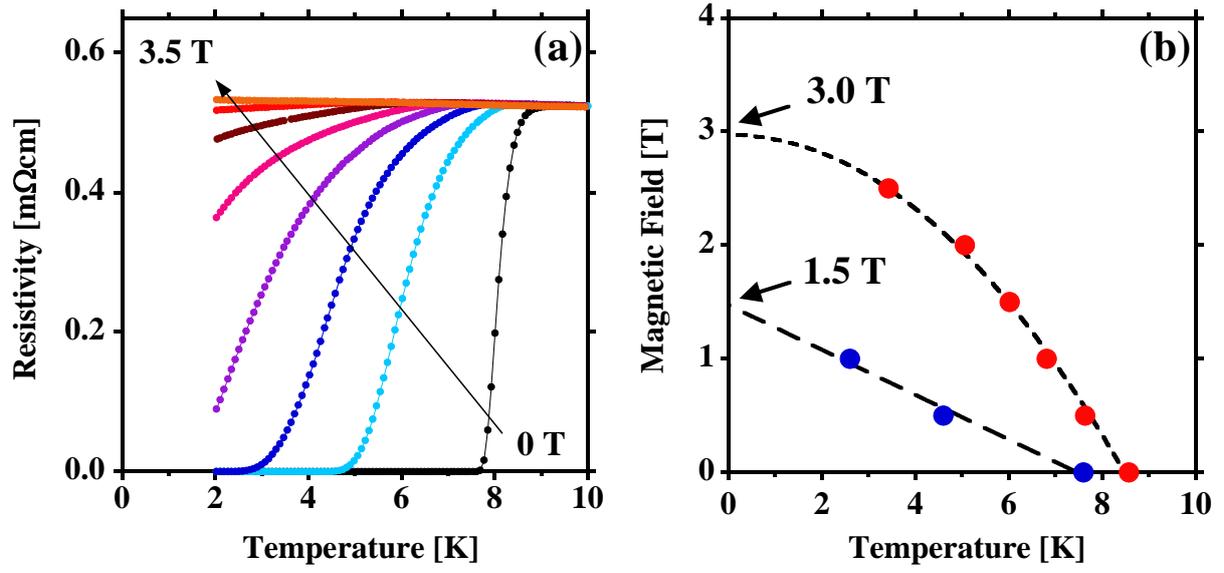

Figure 3